\documentstyle[preprint,eqsecnum,aps]{revtex}
\begin{document}
\draft
\preprint{UT-771, IU-MSTP/19, hep-th/9704062}
\title{Duality in potential curve crossing:\\
Application to quantum coherence}
\author{Kazuo Fujikawa}
\address{%
Department of Physics, University of Tokyo\\
Bunkyo-ku, Tokyo 113, Japan}
\author{Hiroshi Suzuki\cite{email}}
\address{
Department of Physics, Ibaraki University, Mito 310, Japan}
\date{\today}
\maketitle
\begin{abstract}
A field dependent $su(2)$~gauge transformation connects between the
adiabatic and diabatic pictures in the (Landau-Zener-Stueckelberg)
potential curve crossing problem. It is pointed out that weak and
strong potential curve crossing interactions are interchanged under
this transformation, and thus realizing a naive strong and weak
duality. A reliable perturbation theory should thus be formulated in
the both limits of weak and strong interactions. In fact, main
characteristics of the potential crossing phenomena such as the
Landau-Zener formula including its numerical coefficient are
well-described by simple (time-independent) perturbation theory
without referring to Stokes phenomena. We also show that quantum
coherence in a double well potential is generally suppressed by the
effect of potential curve crossing, which is analogous to the effect
of Ohmic dissipation on quantum coherence.
\end{abstract}
\section{Introduction}
\label{sec:one}

The potential curve crossing is related to a wide range of physical
and chemical processes, and the celebrated Landau-Zener
formula~\cite{1,2,3} correctly describes the qualitative features of
those processes~\cite{4,5,6,7,8}. An effort to improve the
Landau-Zener formula and to make it a more quantitative formula is
actively going on even recently~\cite{9,10}. The purpose of the
present paper is to study this old problem on the basis of modern
field theoretical ideas, namely, the duality and gauge transformation.

The adiabatic and diabatic pictures in potential curve crossing
problem are related to each other by a field dependent $su(2)$~gauge
transformation~\cite{5,8}, and we point out that this transformation
leads to an interchange of strong and weak potential curve crossing
interactions, which is analogous to the electric and magnetic duality
in conventional gauge theory~\cite{11}. This strong and weak duality
allows a reliable perturbative treatment of potential curve crossing
phenomena at the both limits of very weak (adiabatic picture) and very
strong (diabatic picture) potential  crossing interactions. Main
features of potential curve crossing phenomena are well-described by
straightforward time-independent perturbation theory combined with the
zeroth order WKB wave functions~\cite{5,8,12}, without referring to
Stokes phenomena; perturbation theory thus becomes more flexible to
cover a wide range of problems.

To illustrate our observation, we first show that simple
time-independent perturbation theory gives an adequate description of
the Landau-Zener formula including numerical coefficients in the both
limits of adiabatic and diabatic pictures. We encounter an interesting
topological object in the present formulation. We then apply our
formulation to the analysis of the effect of potential curve crossing
on quantum coherence in a double-well potential. We show that the
potential crossing generally suppresses quantum coherence, which is
analogous to the effect of Ohmic dissipation on quantum
coherence~\cite{13,14}. To our knowledge, this clear recognition of
duality and the precise criterion for naive perturbation theory
including the analysis of quantum coherence have not been discussed
before.

\section{A model Hamiltonian of potential curve crossing and duality}
\label{sec:two}

To analyze the potential curve crossing, we start with a model
Hamiltonian defined in the so-called diabatic picture~\cite{5,8}
\begin{equation}
   H={1\over 2m}\hat p^2+{V_1(x)+V_2(x)\over2}
     +{V_1(x)-V_2(x)\over2}\sigma_3+{1\over g}\sigma_1
\label{eq:two.one}
\end{equation}
where $\sigma_3$ and~$\sigma_1$ stand for the Pauli matrices. We
assume throughout this article that the potential crossing occurs at
the origin, $V_1(0)=V_2(0)=0$ (see~Fig.~\ref{fig1}). We also take a
convention that the slope of the first potential at the crossing
point is positive, $V_1'(0)>0$. The sign of $V_2'(0)$ may be either
the same as that of $V_1'(0)$ (Section~\ref{sec:three}) or opposite
(Section~\ref{sec:four}).

If one neglects the last term in the above Hamiltonian, one obtains 
the unperturbed Hamiltonian in the diabatic picture
\begin{equation}
   H_0\equiv{1\over 2m}\hat p^2+{V_1(x)+V_2(x)\over2}
   +{V_1(x)-V_2(x)\over2}\sigma_3.
\label{eq:two.two}
\end{equation}
This Hamiltonian~$H_0$ describes two potentials, which are decoupled
from each other. The last term in~(\ref{eq:two.one}),
$H_I\equiv\sigma_1/g$ with a constant~$g$, causes the transition
between these two otherwise independent potential curves. In other
words, if one takes $g\rightarrow{\rm large}$, this case physically
corresponds to a {\it complete\/} potential  crossing from a view
point of {\it adiabatic\/} two-potential crossing~(Fig.~\ref{fig2}).
Namely, $g$~stands for the strength of potential crossing interaction,
and $g\rightarrow{\rm large}$ corresponds to a very strong potential
crossing interaction. On the other hand, if one lets $g$~smaller, the
effects of the last term in~(\ref{eq:two.one}) become substantial and
the Hamiltonian $H_0$~(\ref{eq:two.two}) does not present a sensible
zeroth order Hamiltonian.

To deal with the case of a small~$g$, we perform the non-Abelian
``gauge transformation,''
\begin{equation}
   \Phi(x)=e^{i\theta(x)\sigma_2/2}\Psi(x),\quad
   H'=e^{i\theta(x)\sigma_2/2}He^{-i\theta(x)\sigma_2/2},
\label{eq:two.three}
\end{equation}
where $\sigma_2$~is a Pauli matrix. The Hamiltonian in the new picture
is given by
\begin{eqnarray}
   H'&=&{1\over2m}\left[\hat p
         -{\hbar\over2}\partial_x\theta(x)\sigma_2\right]^2
   +{V_1(x)+V_2(x)\over2}
\nonumber\\
   &&+\left[{V_1(x)-V_2(x)\over2}\cos\theta(x)
                           +{1\over g}\sin\theta(x)\right]\sigma_3
\nonumber\\
   &&+\left[-{V_1(x)-V_2(x)\over2}\sin\theta(x)
                           +{1\over g}\cos\theta(x)\right]\sigma_1.
\label{eq:two.four}
\end{eqnarray}
To eliminate the potential curve mixing, the last term
of~(\ref{eq:two.four}), we choose the gauge parameter~$\theta(x)$
as~\cite{8}
\begin{equation}
   \cot\theta(x)=g{V_1(x)-V_2(x)\over2}\equiv f(x).
\label{eq:two.five}
\end{equation}
We then obtain the Hamiltonian in the {\it adiabatic\/} picture
\begin{equation}
   H'=H_0'+H_I',
\label{eq:two.six}
\end{equation}
where
\begin{equation}
   H_0'\equiv{1\over 2m}\hat p^2+{U_1(x)+U_2(x)\over2}
   +{U_1(x)-U_2(x)\over2}\sigma_3,
\label{eq:two.seven}
\end{equation}
and
\begin{equation}
   H_I'\equiv-{\hbar\over4m}
   \left[\hat p\partial_x\theta(x)
         +\partial_x\theta(x)\hat p\right]\sigma_2
   +{\hbar^2\over8m}\left[\partial_x\theta(x)\right]^2.
\label{eq:two.eight}
\end{equation}
The potential energies in the adiabatic picture are related to those
in the diabatic picture as~(Fig.~\ref{fig2})
\begin{equation}
   U_{1,2}(x)\equiv{V_1(x)+V_2(x)\over2}
   \pm\sqrt{\left[{V_1(x)-V_2(x)\over2}\right]^2+{1\over g^2}}.
\label{eq:two.nine}
\end{equation}
{}From the definition of the gauge parameter in~(\ref{eq:two.five}),
the ``gauge field''~$\partial_x\theta(x)$ is expressed as
\begin{equation}
   \partial_x\theta(x)=-{f'(x)\over1+f(x)^2}.
\label{eq:two.ten}
\end{equation}

The transition from the diabatic picture to the adiabatic picture is
a local gauge transformation, or in the conventional field theoretical
sense it is regarded as a field dependent transformation. The
transformation from one of these two pictures to the other is an
$x$-dependent notion.

In the adiabatic picture, the $\sigma_2$~dependent term in the
interaction~$H_I'$~(\ref{eq:two.eight}) causes the potential 
crossing. If one neglects~$H_I'$, the two potentials characterized by
$U_1(x)$ and~$U_2(x)$ do not mix with each other: Physically, this
means {\it no\/} potential  crossing. This suggests that $H_I'$~is
proportional to the coupling constant~$g$, since a small~$g$
corresponds to {\it weak\/} potential  crossing by definition. This is
in fact the case as is clear from (\ref{eq:two.ten})
and~(\ref{eq:two.five}).

We thus conclude that the two extreme limits of potential crossing
interaction should be reliably handled in perturbation theory; namely,
the strong potential crossing interaction in the {\it diabatic\/}
picture, and the weak potential crossing interaction in the gauge
transformed {\it adiabatic\/} picture. This is analogous to the
electric-magnetic duality in conventional gauge theory~\cite{11}:
The diabatic picture may correspond to the electric picture with a
coupling constant~$e=1/g$, and the adiabatic picture to the magnetic
picture with a coupling constant~$g$.

A general criterion for the validity of perturbation theory in the
adiabatic picture~(\ref{eq:two.six}) is
\begin{equation}
  {\hbar\over2}|\partial_x\theta(x)|\ll|p(x)|,
\label{eq:two.eleven}
\end{equation}
which is expected to be satisfied when the coupling constant~$g$ is
small and the incident particle is sufficiently energetic.

\section{Landau--Zener formula}
\label{sec:three}

As an application of the duality discussed in Section~\ref{sec:two},
we re-examine a perturbative derivation of the Landau-Zener formula in
both of the adiabatic and diabatic pictures~\cite{1,5,8}. For
definiteness, we shall assume $V_1'(0)>V_2'(0)$ as in~Fig.~\ref{fig1}.

Let us start with the adiabatic picture with weak potential crossing
interaction. Since the gauge field generally vanishes,
$\partial_x\theta(x)\rightarrow0$ for $|x|\rightarrow\infty$, we can
define the {\it asymptotic\/} states in terms of the eigenstates
of~$H_0'$~(\ref{eq:two.seven}). We define the initial and final
states $\Phi_i$ and~$\Phi_f$ by
\begin{equation}
   \Phi_i(x)=\pmatrix{\varphi_1(x)\cr 0\cr},\quad
   \Phi_f(x)=\pmatrix{0\cr \varphi_2(x)\cr},
\label{eq:three.one}
\end{equation}
which satisfy
\begin{equation}
   \left[{1\over2m}\hat p^2+U_1(x)\right]\varphi_1(x)
   =E\varphi_1(x),\quad
   \left[{1\over2m}\hat p^2+U_2(x)\right]\varphi_2(x)
   =E\varphi_2(x).
\label{eq:three.two}
\end{equation}
We then obtain the potential curve crossing probability due to the
perturbation~$H_I'$~(\ref{eq:two.eight})
\begin{equation}
   w(i\rightarrow f)
   ={2\pi\over\hbar}|\langle\Phi_f|H_I'|\Phi_i\rangle|^2.
\label{eq:three.three}
\end{equation}
The transition matrix element is given by
\begin{equation}
   \langle\Phi_f|H_I'|\Phi_i\rangle=
   -{\hbar^2\over4m}\int_{-\infty}^\infty dx\,\partial_x\theta(x)
   \left[-\varphi_2'(x)\varphi_1(x)
         +\varphi_2(x)\varphi_1'(x)\right].
\label{eq:three.four}
\end{equation}

To evaluate the matrix element, we use the WKB wave
functions~\cite{4}:
\begin{equation}
   \varphi_1(x)=\cases{
   \displaystyle
   {C_1\over2\sqrt{|p_1(x)|}}
   \exp\left[-{1\over\hbar}\int_{a_1}^xdx\,|p_1(x)|\right]
   &for $x>a_1$,\cr
   \displaystyle
   {C_1\over\sqrt{p_1(x)}}
   \cos\left[{1\over\hbar}\int_x^{a_1}dx\,p_1(x)
   -{\pi\over4}\right]&for $x<a_1$,\cr}
\label{eq:three.five}
\end{equation}
and
\begin{equation}
   \varphi_2(x)=\cases{
   \displaystyle
   {C_2\over2\sqrt{|p_2(x)|}}
   \exp\left[-{1\over\hbar}\int_{a_2}^xdx\,|p_2(x)|\right]
   &for $x>a_2$,\cr
   \displaystyle
   {C_2\over\sqrt{p_2(x)}}
   \cos\left[{1\over\hbar}\int_x^{a_2}dx\,p_2(x)
   -{\pi\over4}\right]&for $x<a_2$,\cr}
\label{eq:three.six}
\end{equation}
where the semi-classical momenta in the adiabatic picture are defined
by
\begin{equation}
   p_{1,2}(x)\equiv\sqrt{2m[E-U_{1,2}(x)]},
\label{eq:three.seven}
\end{equation}
and $a_1$ and~$a_2$ denote the classical turning
points~(Fig.~\ref{fig2}). Within the WKB approximation, we also have
\begin{equation}
   \varphi_1'(x)\simeq\cases{
   \displaystyle
   -{C_1\over2\hbar}\sqrt{|p_1(x)|}
   \exp\left[-{1\over\hbar}\int_{a_1}^xdx\,|p_1(x)|\right]
   &for $x>a_1$,\cr
   \displaystyle
   {C_1\over\hbar}\sqrt{p_1(x)}
   \sin\left[{1\over\hbar}\int_x^{a_1}dx\,p_1(x)
   -{\pi\over4}\right]&for $x<a_1$,\cr}
\label{eq:three.eight}
\end{equation}
and a similar relation for $\varphi_2'(x)$. The normalization
of~$\varphi_1(x)$ is chosen as $C_1=2\sqrt{m}$ to make the
probability flux of the incident wave unity. On the other hand, the
final state wave function in~(\ref{eq:three.three}) has to be
normalized by the delta function with respect to the energy,
$\langle\Phi_2'|\Phi_2\rangle=\delta(E_2'-E_2)$ and this specifies
$C_2=2\sqrt{m}/\sqrt{2\pi\hbar}$.

We estimate the matrix element~(\ref{eq:three.four}) by using the
oscillating parts of the wave functions (\ref{eq:three.five})
and~(\ref{eq:three.six}). This treatment is justified if the
following conditions are satisfied:\\
(i)~$|p(0)|\rightarrow{\rm large}$ and $m\rightarrow{\rm large}$ with
$v=|p(0)|/m$ kept fixed such that non-relativistic treatment is valid
in the physically relevant region.\\
(ii)~$g\rightarrow{\rm small}$, but with
\begin{equation}
   {1\over g}\ll{1\over2m}|p(0)|^2
\label{eq:three.eleven}
\end{equation}
to ensure~(\ref{eq:two.eleven}) and the condition
\begin{equation}
   \beta\ll a,
\label{eq:three.twelve}
\end{equation}
where $a$~is an average turning point and $\beta$~is a typical
geometrical extension of $\partial_x\theta(x)$. If
(\ref{eq:three.twelve})~is satisfied, we can estimate the matrix
element by using the oscillating parts of wave functions only since
$\partial_{x}\theta(x)$ rapidly goes to zero for $|x|\gg\beta$ on the
real axis.

The integral~(\ref{eq:three.four}) is then written as
\begin{eqnarray}
   &&\langle\Phi_f|H_I'|\Phi_i\rangle
   \simeq-{i\hbar C_1C_2\over 8m}\int dx\,\partial_x\theta(x)
  \nonumber\\
   &&\times\biggl\{
   \exp\left[{i\over\hbar}\int_{a_1}^xdx\,p_1(x)
             -{i\over\hbar}\int_{a_2}^xdx\,p_2(x)\right]
     -\exp\left[-{i\over\hbar}\int_{a_1}^xdx\,p_1(x)
             +{i\over\hbar}\int_{a_2}^xdx\,p_2(x)\right]\biggr\},
\label{eq:three.nine}
\end{eqnarray}
where we have set $p_1(x)/p_2(x)=1$ in the prefactors. This is
justified within the saddle point approximation
for~$\hbar\rightarrow0$; this is also justified
if~$\hbar/|p(0)|\ll\beta$, the characteristic length scale of the
present problem, by letting~$p(0)$ large as is specified in~(i).
Therefore we need to evaluate an integral of the form
\begin{equation}
   I\equiv\int_{-\infty}^\infty
   dx\,\partial_x\theta(x)
   \exp\left[{i\over\hbar}\int_{a_1}^xdx\,p_1(x)
             -{i\over\hbar}\int_{a_2}^xdx\,p_2(x)\right].
\label{eq:three.ten}
\end{equation}

We here present an explicit evaluation of~(\ref{eq:three.ten}) for
the linear potential crossing problem, $V_1(x)=V_1'(0)x$ and
$V_2(x)=V_2'(0)x$, on the basis of local data without referring to
Stokes phenomena. For sufficiently large energy,
$E-(U_1+U_2)/2\gg(U_1-U_2)/2$, the difference of momenta can be
approximated as [see~(\ref{eq:two.nine})],
\begin{equation}
   \int_0^xdx\,[p_1(x)-p_2(x)]
   \simeq-\int_0^xdx\,{2\over v(x)g\beta}\,\sqrt{x^2+\beta^2}
\label{eq:three.thirteen}
\end{equation}
where we used 
\begin{equation}
   f(x)=g{V_1'(0)-V_2'(0)\over2}x
   \equiv{x\over\beta},\quad
   v(x)\equiv{1\over m}
   \sqrt{2m\left[E-{U_1(x)+U_2(x)\over2}\right]}
\label{eq:three.fourteen}
\end{equation}
and $v(x)$~is approximated to be a constant $v=v(0)$ in the
following. We also have from~(\ref{eq:two.ten})
\begin{equation}
   \partial_x\theta(x)=-{\beta\over x^2+\beta^2}
\label{eq:three.fifteen}
\end{equation}
and thus
\begin{eqnarray}
   I&\simeq&
   -\exp\left[{i\over\hbar}\int_{a_1}^0dx\,p_1(x)
         -{i\over\hbar}\int_{a_2}^0dx\,p_2(x)\right]
   \int_{-\infty}^\infty dx\,{\beta\over x^2+\beta^2}
   \exp\left(-{2i\over\hbar vg\beta}
   \int_0^xdx\,\sqrt{x^2+\beta^2}\right)
\nonumber\\
   &=&
   -\exp\left[{i\over\hbar}\int_{a_1}^0dx\,p_1(x)
         -{i\over\hbar}\int_{a_2}^0dx\,p_2(x)\right]
   \int_{-\infty}^\infty dx\,{1\over x^2+1}
   \exp\left(-i\alpha\int_0^xdx\,\sqrt{x^2+1}\right)
\nonumber\\ 
   &=&
   -\exp\left[{i\over\hbar}\int_{a_1}^0dx\,p_1(x)
         -{i\over\hbar}\int_{a_2}^0dx\,p_2(x)\right]
   \int_{-\infty}^\infty dx\,\exp[-i\alpha F(x)]
\label{eq:three.sixteen}
\end{eqnarray}
where
\begin{equation}
   F(x)\equiv\int_0^xdx\,\sqrt{x^2+1}+{1\over i\alpha}\ln(x^2+1),
   \quad\alpha\equiv{2\beta\over\hbar vg}
   ={4\over\hbar vg^2[V_1'(0)-V_2'(0)]}>0. 
\label{eq:three.seventeen}
\end{equation}
We evaluate the integral~(\ref{eq:three.sixteen}) by a saddle point
approximation with respect to~$\alpha$. We thus seek the saddle point
\begin{equation}
   F'(x)=\sqrt{x^2+1}+{1\over i\alpha}{2x\over x^2+1}=0,
\label{eq:three.eighteen}
\end{equation}
which is located between the real axis and the pole
positions~$x=\pm i$ of~$\partial_x\theta(\beta x)$ so that we can
smoothly deform the integration contour; these poles also coincide
with the complex potential crossing points. If one sets~$x=iy$
in~(\ref{eq:three.eighteen}) for $-1<y<1$, one has
\begin{equation}
   \sqrt{1-y^2}=-{1\over\alpha}{2y\over1-y^2}
\label{eq:three.nineteen}
\end{equation}
which has a {\it unique\/} solution
\begin{equation}
   x_s=iy_s\simeq-i+{i\over2}\left({2\over\alpha}\right)^{2/3}
\label{eq:three.twenty}
\end{equation}
for {\em large}~$\alpha$. (The complex conjugate of~$x_s$ is
located in the second Riemann sheet.)  For this value of the saddle
point
\begin{eqnarray}
   &&F(x_s)=\int_0^{x_s}dx\,\sqrt{x^2+1}
     +{1\over i\alpha}\ln\left({2\over\alpha}\right)^{2/3}
   \simeq-{\pi i\over4}+{2\over3}{i\over\alpha}
   +{1\over i\alpha}\ln\left({2\over\alpha}\right)^{2/3},
\nonumber\\
   &&F''(x_s)\simeq-3i\left({\alpha\over2}\right)^{1/3}.
\label{eq:three.twentyone}
\end{eqnarray}
We thus have a Gaussian integral which decreases in the direction
{\it parallel\/} to the real axis
\begin{eqnarray}
   I&\simeq&
   -\left({\alpha\over2}\right)^{2/3}e^{2/3}
   \exp\left[{i\over\hbar}\int_{a_1}^0dx\,p_1(x)
             -{i\over\hbar}\int_{a_2}^0dx\,p_2(x)\right]
\nonumber\\
   &&\qquad\qquad\qquad\times
   \int_{-\infty}^\infty dx\,
   \exp\left[-3\left(\alpha\over2\right)^{4/3}(x - x_s)^2\right]
   \exp\left(-{\pi\alpha\over4}\right)
\nonumber\\
   &=&
   -\sqrt{\pi\over3}e^{2/3}
   \exp\left[{i\over\hbar}\int_{a_1}^0dx\,p_1(x)
         -{i\over\hbar}\int_{a_2}^0dx\,p_2(x)\right]
   \exp\left(-{\pi\alpha\over4}\right).
\label{eq:three.twentytwo}
\end{eqnarray}
{}From~(\ref{eq:three.nine}) we obtain
\begin{eqnarray}
   \langle\Phi_f|H_I'|\Phi_i\rangle
   &\simeq&
   -\sqrt{\pi\over3}e^{2/3}\sqrt{\hbar\over2\pi}
   \sin\left\{{1\over\hbar}
   \left[\int_{a_1}^{0}dx\,p_1(x)-\int_{a_2}^{0}dx\,p_2(x)\right]
   \right\}
\nonumber\\
   &&\qquad\qquad\qquad\times
   \exp\left\{-{\pi\over\hbar vg^2[V_1'(0)-V_2'(0)]}
   \right\}.
\label{eq:three.twentythree}
\end{eqnarray}
It is interesting that the numerical value of the coefficient of the
above expression, $\sqrt{\pi}e^{2/3}/\sqrt{3}=1.99317$, is very close
to the canonical value~$2$~\cite{4}, and we replace it by~$2$ in the
following. We thus have the transition probability
from~(\ref{eq:three.three})
\begin{eqnarray}
   w(i\rightarrow f)&\simeq&
   4\sin^2\left\{{1\over\hbar}\,
   \left[\int_{a_1}^0dx\,p_1(x)-\int_{a_2}^0dx\,p_2(x)\right]\right\}
   \exp\left\{-{2\pi\over\hbar vg^2[V_1'(0)-V_2'(0)]}\right\}
\nonumber\\
    &\simeq&2\exp\left\{-{2\pi\over\hbar vg^2[V_1'(0)-V_2'(0)]}
     \right\}
\label{eq:three.twentyfour}
\end{eqnarray}
where we replaced the square of sine function by its average~$1/2$
in the final expression. We emphasize that the numerical coefficient
of~$w(i\rightarrow f)$ is fixed by perturbation theory and the local
data without referring to global Stokes phenomena; this is
satisfactory since linear potential crossing is a locally valid
idealization.

We interpret that $w(i\rightarrow f)$~in~(\ref{eq:three.twentyfour})
expresses {\it twice\/} the non-adiabatic transition probability.
Notice that our initial state wave function contains the reflection
wave as well as the incident wave. Therefore the transition
probability per {\it one\/} crossing is given by the half
of~(\ref{eq:three.twentyfour}),
\begin{equation}
   P(1\rightarrow2)\simeq
   \exp\left\{-{2\pi\over\hbar vg^2[V_1'(0)-V_2'(0)]}\right\},
\label{eq:three.twentyseven}
\end{equation}
which is the celebrated Landau-Zener formula~\cite{4,5}. Our
perturbative derivation presented here is conceptually much simpler
than the past works~\cite{1,4,5,8,12}, and it should be useful for a
pedagogical purpose also.

It is interesting to study the same problem in the diabatic picture
in~Fig.~\ref{fig1} with~$H_I=\sigma_1/g$ for large~$g$. We first note
that the initial state in the adiabatic picture asymptotically
corresponds to the eigenfunction of~$V_2(x)$ and the final state
corresponds to~$V_1(x)$, under the present setup $V_1'(0)>V_2'(0)$.
Therefore the initial and final zeroth order wave functions in the
diabatic picture are taken as
\begin{equation}
   \Psi_i(x)=\pmatrix{0\cr\psi_2(x)\cr},\quad
   \Psi_f(x)=\pmatrix{\psi_1(x)\cr0\cr}.
\label{eq:three.twentyeight}
\end{equation}
In the WKB approximation,
\begin{equation}
   \psi_1(x)=\cases{
   \displaystyle
   {C_1\over2\sqrt{|p_1(x)|}}
   \exp\left[-{1\over\hbar}\int_{a_1}^xdx\,|p_1(x)|\right]
   &for $x>a_1$,\cr
   \displaystyle
   {C_1\over\sqrt{p_1(x)}}
   \cos\left[{1\over\hbar}\int_x^{a_1}dx\,p_1(x)
   -{\pi\over4}\right]&for $x<a_1$,\cr}
\label{eq:three.twentynine}
\end{equation}
and
\begin{equation}
   \psi_2(x)=\cases{
   \displaystyle
   {C_2\over2\sqrt{|p_2(x)|}}
   \exp\left[-{1\over\hbar}\int_{a_2}^xdx\,|p_2(x)|\right]
   &for $x>a_2$,\cr
   \displaystyle
   {C_2\over\sqrt{p_2(x)}}
   \cos\left[{1\over\hbar}\int_x^{a_2}dx\,p_2(x)
   -{\pi\over4}\right]&for $x<a_2$.\cr}
\label{eq:three.thirty}
\end{equation}
In the above expressions, semi-classical momenta are defined by
\begin{equation}
      p_{1,2}(x)\equiv\sqrt{2m[E-V_{1,2}(x)]}.
\label{eq:three.thirtyone}
\end{equation}
The transition probability in the diabatic picture is then given by
\begin{equation}
   w(i\rightarrow f)
   ={2\pi\over\hbar}|\langle\Psi_f|H_I|\Psi_i\rangle|^2.
\label{eq:three.thirtytwo}
\end{equation}
The evaluation of the matrix element in~(\ref{eq:three.thirtytwo})
is the standard one described in the textbook of Landau and
Lifshitz~\cite{4}, for example. The saddle point for~$\hbar\ll1$ is
located at the origin~$x_s=0$ and we have an integral, for example,
\begin{eqnarray}
   &&\int_{-\infty}^\infty dx\,{1\over\sqrt{p_1(x)p_2(x)}}
   \exp\left[{i\over\hbar}\int_{a_1}^xdx\,p_1(x)
             -{i\over\hbar}\int_{a_2}^xdx\,p_2(x)\right]
\nonumber\\
   &&\simeq{\sqrt{2\pi\hbar}
      \over\sqrt{m[V_1'(0)-V_2'(0)]}(2mE)^{1/4}}
   \exp\left[{i\over\hbar}\int_{a_1}^0dx\,p_1(x)
             -{i\over\hbar}\int_{a_2}^0dx\,p_2(x)
             -{\pi i\over4}\right].
\label{eq:three.thirtythree}
\end{eqnarray}
Under this approximation, we have for $E>0$,
\begin{eqnarray}
   w(i\rightarrow f)
   &\simeq&
   {8\pi\over\hbar v(0)g^2[V_1'(0)-V_2'(0)]}
   \cos^2\left[{1\over\hbar}\int_0^{a_2}dx\,p_2(x)
               -{1\over\hbar}\int_0^{a_1}dx\,p_1(x)
               -{\pi\over4}\right]
\nonumber\\
   &\simeq&{4\pi\over\hbar vg^2[V_1'(0)-V_2'(0)]}.
\label{eq:three.thirtyfour}
\end{eqnarray}
[$v(0)$ is the velocity at the crossing point, $v(0)=\sqrt{2E/m}$.]
For $E<0$, one can verify that there is no saddle point and
therefore,
\begin{equation}
   w(i\rightarrow f)\simeq0.
\label{eq:three.thirtyfive}
\end{equation}
We again interpret~(\ref{eq:three.thirtyfour}) as twice the potential
crossing probability because our initial state wave function contains
the reflection wave as well as the incident wave. The transition
probability per one potential crossing is given by the half
of~(\ref{eq:three.thirtyfour}). 

A simple {\it interpolating\/} formula, which
reproduces~(\ref{eq:three.twentyfour}) in the weak coupling limit and
(\ref{eq:three.thirtyfour}) in the strong coupling limit, is given by
\begin{equation}
   w(i\rightarrow f)\simeq
   2\exp\left\{-{2\pi\over\hbar vg^2[V_1'(0)-V_2'(0)]}\right\}
   \left(1-
    \exp\left\{-{2\pi\over\hbar vg^2[V_1'(0)-V_2'(0)]}\right\}
   \right)
\label{eq:three.thirtysix}
\end{equation}
This expression is also consistent with the (semi-classical)
conservation of probability~\cite{4}. A more rigorous justification 
of~(\ref{eq:three.thirtysix}) is facilitated by combining the analysis
of Stokes phenomena and the conservation of probability~\cite{6}.

Motivated by duality, we re-examined the perturbative derivation of
the Landau-Zener formula, and we re-derived the
formula~(\ref{eq:three.twentyfour}) including its numerical
coefficient on the basis of perturbation theory. However, our final
formula~(\ref{eq:three.twentyfour}) in the adiabatic picture does not
contain the coupling constant as a prefactor. This is related to an
interesting topological object in the present formulation. From the
definition of~(\ref{eq:two.ten}), the ``gauge field'' satisfies the
relation
\begin{eqnarray}
   \int_{-\infty}^\infty dx\,\partial_x\theta(x)
   &=&\theta(\infty)-\theta(-\infty)
\nonumber\\  
   &=&-\pi,
\label{eq:three.thirtyseven}
\end{eqnarray}
which is {\it independent\/} of the coupling constant~$g$; we assume 
$f(x)\rightarrow\pm\infty$ for $x\rightarrow\pm\infty$, respectively.
Because of this kink-like topological behavior of~$\theta(x)$, the
coupling constant does not appear as a prefactor of the matrix
element in perturbation theory if the wave functions spread over the
range which well covers the geometrical size of~$\partial_x\theta(x)$.
The precise criterion of the validity of perturbation theory is thus
given by~(\ref{eq:two.eleven}): This condition is in fact satisfied if
the conditions (\ref{eq:three.eleven})--(\ref{eq:three.twelve}) are
satisfied. For small values of~$x$, the small coupling~$g$ helps to
satisfy~(\ref{eq:two.eleven}). Even for the values of~$x$ near the
average turning point~$a$, we have
\begin{equation}
  {\hbar\over2}|\partial_x\theta(a)|
   \simeq{1\over2}\left({\beta\over a}\right){\hbar\over a}\ll
   {\hbar\over a}\simeq|p(a)|,
\label{eq:three.thirtyeight}
\end{equation}
where $\beta$~stands for the typical geometrical size
of~$\partial_x\theta(x)$. The estimate in the left hand side is based
on linear potentials~(\ref{eq:three.fifteen}), but we expect that the
condition is satisfied for more general potentials as well.

To conclude this section, we clarified the basic mechanism why the
prefactor of the Landau-Zener formula~(\ref{eq:three.twentyseven})
should come out to be very close to unity in time-independent
perturbation theory.

\section{Level crossing and quantum coherence}
\label{sec:four}

The effects of dissipative interactions on macroscopic quantum
tunneling have been extensively analyzed in the path integral
formalism~\cite{13} and also in the canonical (field theoretical)
formalism~\cite{14}. It is generally accepted that the Ohmic
dissipation suppresses the macroscopic quantum coherence; in fact,
an attractive idea of a dissipative phase transition has been
suggested~\cite{13}.

It is plausible that the effects of potential curve crossing with
nearby potentials influence the quantum coherence of the two
degenerate ground states. We here analyze the general features of the
effects of potential crossing on quantum coherence on the basis of
time-independent perturbation theory, which was confirmed in the
preceding section to be reliable in the both limits of weak and strong
potential curve crossing interaction. We assume
$V_2(x)=V_1(-x)$ and $V_2'(0)<0$.

\noindent
We thus analyze this problem from two different view points.

\subsection{Strong potential curve crossing interaction $g\gg1$
(diabatic picture)}

We start with the diabatic picture in~Fig.~\ref{fig4} and incorporate
the effects of the potential crossing interaction~$\sigma_{1}/g$.
Namely, we (approximately) diagonalize the total Hamiltonian in the
diabatic picture. By treating the last term of~(\ref{eq:two.one}) as
a perturbation for~$g\gg1$, we obtain the energy eigenvalues of the
two lowest lying states as (with~$E_0$ the zeroth order degenerate
energy eigenvalue)
\begin{equation}
   E_{\pm}^{(1)}=E_0\mp{1\over g}\int_{-\infty}^\infty dx\,
   \psi_L(x)\psi_R(x),
\label{eq:five.one}
\end{equation}
with the corresponding eigenfunctions
\begin{eqnarray}
   &&\Psi_+(x)
   \simeq{1\over\sqrt{2}}\pmatrix{\psi_L(x)\cr-\psi_R(x)\cr}
   \equiv{1\over\sqrt{2}}[e_+\psi_L(x)-e_-\psi_R(x)],
\nonumber\\
   &&\Psi_-(x)
   \simeq{1\over\sqrt{2}}\pmatrix{\psi_L(x)\cr\psi_R(x)\cr}
   \equiv{1\over\sqrt{2}}\bigl[e_+\psi_L(x)+e_-\psi_R(x)\bigr].
\label{eq:five.two}
\end{eqnarray}
Namely, both of these two states~$\Psi_\pm(x)$ choose
\begin{eqnarray}
   &&\hbox{$e_+\psi_L(x)$ in the region of left valley},
\nonumber\\
   &&\hbox{$e_-\psi_R(x)$ in the region of right valley}.
\label{eq:five.three}
\end{eqnarray}
The ``spin'' eigenstates~$e_\pm$ and the ``space'' eigenstates
$\psi_L(x)$ or~$\psi_R(x)$ are strongly correlated. Note that one can
conveniently think of the two potentials as spin-up and spin-down
states. Originally, before we incorporate the
perturbation~$\sigma_1/g$ in~(\ref{eq:five.one}), we have
energetically degenerate two states with energy~$E_0$ defined by
\begin{equation}
   \Psi_L(x)=\pmatrix{\psi_L(x)\cr0\cr}\equiv e_+\psi_L(x),
\label{eq:five.four}
\end{equation}
which is always a spin-up state, and 
\begin{equation}
   \Psi_R(x)=\pmatrix{0\cr\psi_R(x)\cr}\equiv e_-\psi_R(x),
\label{eq:five.five}
\end{equation}
which is always a spin-down state.

In the strong potential crossing limit, $g\rightarrow\infty$, we have
{\it no effect\/} of the conventional quantum tunneling in addition
to the mixing in~(\ref{eq:five.two}), since the conventional
tunneling goes through the interaction~$\sigma_1/g$
in~(\ref{eq:two.one}), and thus higher order effects in~$1/g$.

We can find the $O(1/g)$~energy splitting in the WKB approximation.
The left wave function~$\psi_L(x)$ is given
by~(\ref{eq:three.twentynine}) with~$C_1=\sqrt{2m\omega/\pi}$
(For a sufficiently low energy state such as the ground state,
$\omega$~can be regarded as the curvature of the potential energy
at the minimum.) The right wave function~$\psi_R(x)$ is given
by
\begin{equation}
   \psi_R(x)=\cases{
   \displaystyle
   {C_2\over2\sqrt{|p_2(x)|}}
   \exp\left[-{1\over\hbar}\int_x^{a_2}dx\,|p_2(x)|\right]
   &for $x<a_2$,\cr
   \displaystyle
   {C_2\over\sqrt{p_2(x)}}
   \cos\left[{1\over\hbar}\int_{a_2}^xdx\,p_2(x)
   -{\pi\over4}\right]&for $x>a_2$.\cr}\nonumber
\end{equation}
 with~$C_2=\sqrt{2m\omega/\pi}$. In the saddle
point approximation, we have the overlap integral
\begin{eqnarray}
   &&\int dx\,{1\over\sqrt{|p_1(x)||p_2(x)|}}
   \exp\left[-{1\over\hbar}\int_{a_1}^xdx\,|p_1(x)|
             +{1\over\hbar}\int_{a_2}^xdx\,|p_2(x)|\right]
\nonumber\\
   &&\simeq{\sqrt{2\pi\hbar}
      \over\sqrt{m[V_1'(0)-V_2'(0)]}(-2mE)^{1/4}}
   \exp\left[-{1\over\hbar}\int_{a_1}^0dx\,|p_1(x)|
             +{1\over\hbar}\int_{a_2}^0dx\,|p_2(x)|\right],
\label{eq:five.six}
\end{eqnarray}
which yields the energy splitting,
\begin{eqnarray}
   &&E_-^{(1)}-E_+^{(1)}
\nonumber\\
   &&\simeq
   {\sqrt{2\hbar}\over g\sqrt{\pi[V_1'(0)-V_2'(0)]}}
   \left({m\over-2E_0}\right)^{1/4}
   \exp\left\{-{1\over\hbar}
   \int_{a_1}^{a_2}dx\,\sqrt{2m[\overline V(x)-E_0]}\right\}.
\label{eq:five.seven}
\end{eqnarray}
In the above expression, we have defined~$\overline V(x)$ as the
following ``$\Lambda$-shape'' potential:
\begin{equation}
   \overline V(x)=\cases{V_1(x)&for $x<0$\cr
               V_2(x)&for $x>0$.\cr}
\label{eq:five.eight}
\end{equation}
We recall that, when the upper adiabatic potential~$U_1(x)$ is absent,
the standard WKB formula gives the energy splitting of lowest
levels~\cite{4},
\begin{equation}
   E_--E_+\simeq
   {\hbar\omega\over\pi}
   \exp\left\{-{1\over\hbar}
   \int_{-a}^adx\,\sqrt{2m[U_2(x)-E_0]}\right\}.
\label{eq:five.nine}
\end{equation}
Comparing this with~(\ref{eq:five.seven}), we find that the quantum
coherence is actually suppressed by two elements; the overall
small coefficient~$1/g\ll1$ and the exponential suppression factor;
$\overline V(x)$~is always larger than the lower potential,
$\overline V(x)>U_2(x)$.

\subsection{Weak potential curve crossing interaction $g\ll1$
(adiabatic picture)}

In this case, we can use the gauge transformed adiabatic picture
defined by~(\ref{eq:two.six}). In this scheme, the low lying
eigenstates for~$H_0'$ are described by the nearly degenerate two
states in the lower potential curve~$U_2(x)$. See~Fig.~\ref{fig5}.
For a small~$g$, the two potential curves $U_1(x)$ and~$U_2(x)$ are
widely separated. We thus take the ground states of the spin-down
sector (i.e., the nearly degenerate two ground states of the lower
potential curve),
\begin{equation}
   \Phi_+(x)=\pmatrix{0\cr\varphi_{2+}(x)\cr},
   \quad
   \Phi_-(x)=\pmatrix{0\cr\varphi_{2-}(x)\cr}
\label{eq:five.ten}
\end{equation}
as the zeroth order approximation of the lowest lying states. These
two states correspond to the conventional symmetric and anti-symmetric
tunneling eigenstates of the double well potential~$U_2(x)$, whose
energy splitting gives the {\it order parameter\/} of the quantum
coherence~\cite{13}. To the second order of the gauge
field~$\partial_x\theta(x)$, the energy eigenvalue is perturbed to
\begin{eqnarray}
   E_\pm^{(2)}&=&
   E_\pm+\langle\Phi_\pm|H_I'|\Phi_\pm\rangle
   -\sum_n{|\langle\Phi_{1,n}|H_I'|\Phi_\pm\rangle|^2
           \over E_{1,n}-E_\pm}
\nonumber\\
   &=&E_\pm+{\hbar^2\over8m}\int dx\,
      [\partial_x\theta(x)]^2\varphi_{2\pm}(x)^2
\nonumber\\
   &&-{\hbar^2\over16m^2}\sum_n{1\over E_{1,n}-E_\pm}
   \left|\int dx\,\varphi_{1,n}(x)
   \left[\hat p\partial_x\theta(x)
         +\partial_x\theta(x)\hat p\right]\varphi_{2\pm}(x)\right|^2,
\label{eq:five.eleven}
\end{eqnarray}
where $\Phi_{1,n}(x)$ [and $\varphi_{1,n}(x)$] is the $n$th energy
eigenstate of the upper potential~$U_1(x)$.

It is generally difficult to conclude the suppression or enhancement
of the quantum coherence solely from the perturbation
formula~(\ref{eq:five.eleven}). The reason is that the effect of
intermediate state sum can depend on the details of the upper
potential curve. Nevertheless, when the well of the upper potential is
narrow enough, we can show that the quantum coherence is always
suppressed.
 
We first assume that the lowest lying parity even and parity odd
states are given respectively by a linear combination of the ground
states in left and right wells:
\begin{equation}
   \varphi_{2\pm}(x)={1\over\sqrt{2}}
   [\varphi_{2L}(x)\pm\varphi_{2R}(x)],
\label{eq:five.twelve}
\end{equation}
where in the WKB approximation in the classically forbidden region,
\begin{equation}
   \varphi_{2L}(x)=\varphi_{2R}(-x)=
   {C_2\over2\sqrt{|p_2(x)|}}
   \exp\left[-{1\over\hbar}\int^xdx\,|p_2(x)|\right].
\label{eq:five.thirteen}
\end{equation} 
The normalization constant is given by~$C_2=\sqrt{2m\omega/\pi}$.

We next assume that the upper potential~$U_1(x)$ is narrow enough
and can be well approximated by a square well potential with the
width~$a$. The eigenfunctions in the upper potential are therefore
given by
\begin{equation}
   \varphi_{1,n}(x)=\sqrt{2\over a}
   \sin\left[{n\pi\over a}(x+{a\over2})\right],\quad
   n=1,2,\cdots.
\label{eq:five.fourteen}
\end{equation}
Moreover, since the gauge field~$\partial_x\theta(x)$ defined
in~(\ref{eq:two.ten}) is non-zero only within this narrow region,
we may make the following replacements in the
integral~(\ref{eq:five.eleven}):
\begin{eqnarray}
   &&\varphi_{2+}(x)\rightarrow\varphi_{2+}(0),\quad
   \varphi_{2-}(x)\rightarrow0,
\nonumber\\
   &&\hat p\,\varphi_{2+}(x)\rightarrow0,\quad
   \hat p\,\varphi_{2-}(x)\rightarrow i|p_2(0)|\varphi_{2+}(0),
\label{eq:five.fifteen}
\end{eqnarray} 
and
\begin{equation}
   \hat p\,\varphi_{1,n}(x)\rightarrow
   -i|p_{1,n}|\sqrt{2\over a}
   \cos\left[{n\pi\over a}\left(x+{a\over2}\right)\right],
   \quad |p_{1,n}|\equiv\hbar{n\pi\over a}.
\label{eq:five.sixteen}
\end{equation}

Under these assumptions, the energy shift~(\ref{eq:five.eleven}) is
estimated for the parity even state,
\begin{eqnarray}
   E_+^{(2)}&=&
   E_++{\hbar^2\over8m}\int dx\,
   [\partial_x\theta(x)]^2\varphi_{2+}(0)^2
\nonumber\\
   &&-{\hbar^2\over16m^2}\sum_{n=1}^\infty
   {|p_{1,n}|^2\over E_{1,n}-E_+}
   \left|\int dx\,\sqrt{2\over a}
   \cos\left[{n\pi\over a}\left(x+{a\over2}\right)\right]
   \partial_x\theta(x)\right|^2\varphi_{2+}(0)^2.
\label{eq:five.seventeen}
\end{eqnarray}
By noting $|p_{1,n}|^2=2m[E_{1,n}-U_1(0)]$ and $U_1(0)>E_+$,
we see that
\begin{equation}
   {\hbar^2\over16m^2}{|p_{1,n}|^2\over E_{1,n}-E_+}=
   {\hbar^2\over8m}{E_{1,n}-U_1(0)\over E_{1,n}-E_+}<
   {\hbar^2\over8m},
\label{eq:five.twenty}
\end{equation} 
and thus
\begin{eqnarray}
   E_+^{(2)}&>&
   E_++{\hbar^2\over8m}\int dx\,
   [\partial_x\theta(x)]^2\varphi_{2+}(0)^2
\nonumber\\
   &&-{\hbar^2\over8m}\sum_{n=0}^\infty
   \left|\int dx\,\sqrt{2\over a}
   \cos\left[{n\pi\over a}\left(x+{a\over2}\right)\right]
   \partial_x\theta(x)\right|^2\varphi_{2+}(0)^2
\nonumber\\
   &=&E_+,
\label{eq:five.twentyone}
\end{eqnarray}
where we have added $n=0$~mode to make the cosine functions a complete
set. Therefore the perturbation always increases the ground state
energy.

On the other hand, eq.~(\ref{eq:five.eleven}) gives for the parity
odd state,
\begin{equation}
   E_-^{(2)}=E_-
   -{\hbar^2\over16m^2}\sum_{n=1}^\infty
   {|p_2(0)|^2\over E_{1,n}-E_-}
   \left|\int dx\,
   \varphi_{1,n}(x)\partial_x\theta(x)\right|^2\varphi_{2+}(0)^2.
\label{eq:five.eighteen}
\end{equation}
Since the second term of~(\ref{eq:five.eighteen}) is negative
definite, we see that the perturbation always lowers the odd state
energy,
\begin{equation}
   E_-^{(2)}<E_-.
\label{eq:five.nineteen}
\end{equation}

The perturbative corrections in (\ref{eq:five.seventeen})
and~(\ref{eq:five.eighteen}) are proportional to the wave
function at the origin, which satisfies
$\varphi_{2+}(0)^2\simeq m(E_--E_+)/[\hbar|p_2(0)|]$~\cite{4}, and
therefore the correction to the energy splitting itself is
proportional to the zeroth order energy splitting, $E_--E_+$. Since
the correction cannot excess the zeroth order value in a reliable
region of perturbation theory, we should have $E_-^{(2)}-E_+^{(2)}>0$.
By combining (\ref{eq:five.twentyone}) and~(\ref{eq:five.nineteen}),
therefore we find
\begin{equation}
   0<E_-^{(2)}-E_+^{(2)}<E_--E_+,
\label{eq:five.twentytwo}
\end{equation}
which shows that the quantum coherence is suppressed.

Although we derived the suppression of quantum coherence for the very
special configuration in the weak coupling adiabatic picture, we
expect that this suppression of quantum coherence is more general
at least for small~$g$. This is based on the following physical
picture: When a particle tunnels the barrier in the lower potential
curve, it has a small probability to cross to the upper potential
curve. This potential curve crossing means that the particle enters
deep inside the tunneling region of the upper potential curve and
repelled back by the upper potential (as is expected in the diabatic
picture), which in turn leads to the suppression of quantum tunneling
of the particle. This argument leads to the general suppression of
quantum coherence by potential crossing. This is also consistent with
the suppression of barrier transmission in the scattering process by
the non-adiabatic potential crossing effect~\cite{8}.

Our explicit analyses both in the weak and strong potential crossing
interactions suggest the general suppression of quantum coherence by
potential crossing, which is analogous to the effect of Ohmic
dissipation on quantum coherence~\cite{13,14}. In this connection, we
may note that the formula~(\ref{eq:five.eleven}) and the Ohmic
dissipation in the Caldeira-Legget model~\cite{13} both correspond to
the dipole approximation. However, an analysis of the basically
non-perturbative tunneling effect in the second order perturbation
theory needs a great care; for this reason, we were able to explicitly
analyze the very specific case in 
(\ref{eq:five.fourteen})--(\ref{eq:five.sixteen}). This suppression
phenomenon of quantum coherence may become important in the future
when one takes the effects of the environment into account in the
analysis of potential curve crossing in physical and chemical
processes.

\section{Discussion and Conclusion}

Motivated by the presence of interesting weak and strong duality in
the model Hamiltonian~(\ref{eq:two.one}) of potential curve crossing,
we analyzed a perturbative approach to potential crossing phenomena.
We have shown that straightforward perturbation theory combined with
the zeroth order WKB wave functions provides a reliable description
of general potential crossing phenomena. Our analysis is based on the
local data as much as possible without referring to global Stokes
phenomena~\cite{12}. Formulated in this manner, perturbation theory
becomes more flexible to cover a wide range of problems. From a
perturbative view point, the treatment of the Landau-Zener formula
in the adiabatic picture is most complicated since it does not contain
the coupling constant in the prefactor. We pointed out that this
absence of the coupling constant in the prefactor is related to the
presence of an interesting kink-like topological object in the present
formulation of adiabatic picture. The absence of the coupling constant
in the prefactor is thus perfectly consistent with perturbation
theory, provided that a more precise criterion of perturbation
theory~(\ref{eq:two.eleven}) is satisfied. In effect, we explained
why the prefactor of the Landau-Zener formula should come out to be 
very close to unity in perturbation theory.

The suppression of quantum coherence by potential curve crossing,
which to our knowledge has not been discussed before in this context,
has also been neatly formulated in our treatment, in the both limits
of very strong and very weak potential crossing interactions.

{}From a view point of general gauge theory, it is not unlikely that
the electric-magnetic duality in conventional gauge theory is also
related to some generalized form of potential crossing in the
so-called moduli space~\cite{11}. We hope that our work may turn out
to be relevant from this view point also.

Finally, we gratefully acknowledge H. Nakamura for critical comments
which greatly helped improve the present work.



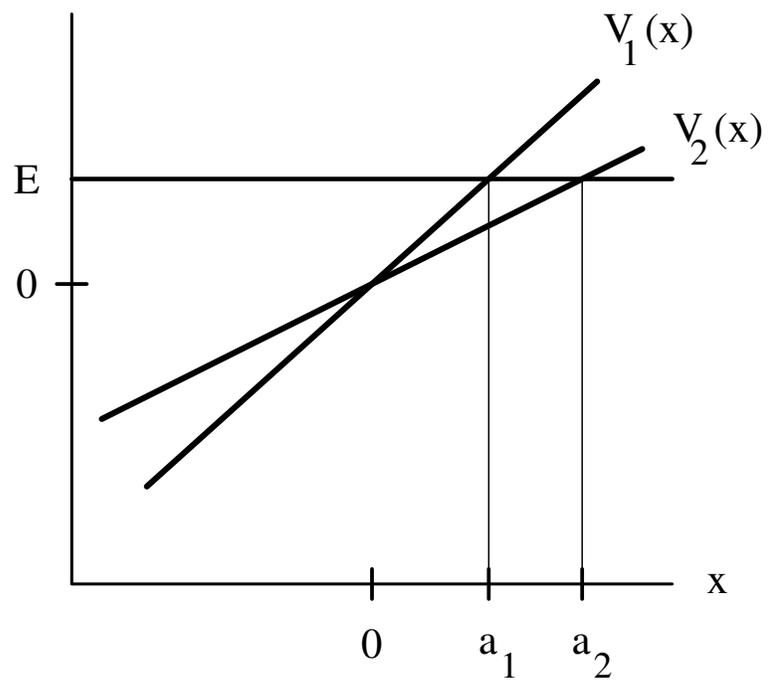
\begin{figure}
\caption{Landau--Zener process in the diabatic picture.}
\label{fig1}
\end{figure}
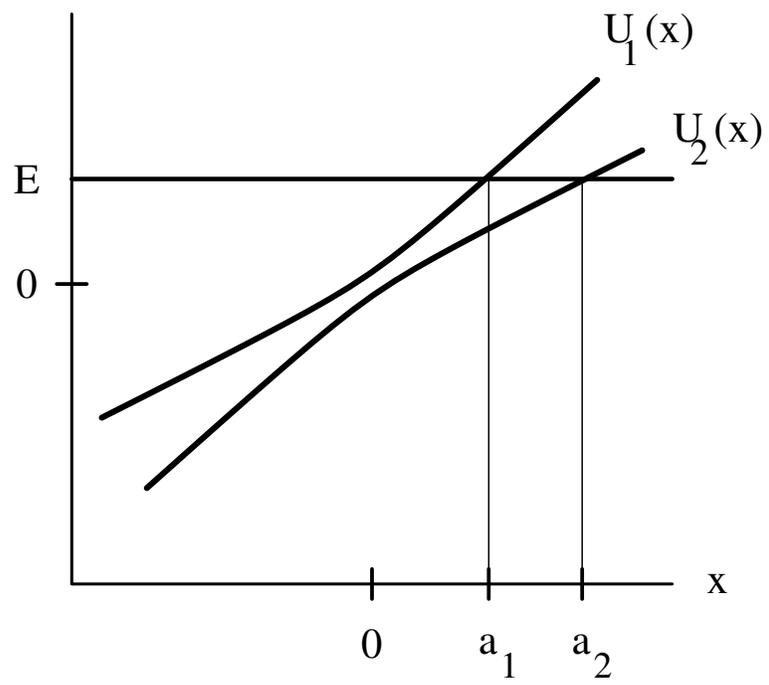
\begin{figure}
\caption{Landau--Zener process in the adiabatic picture.}
\label{fig2}
\end{figure}
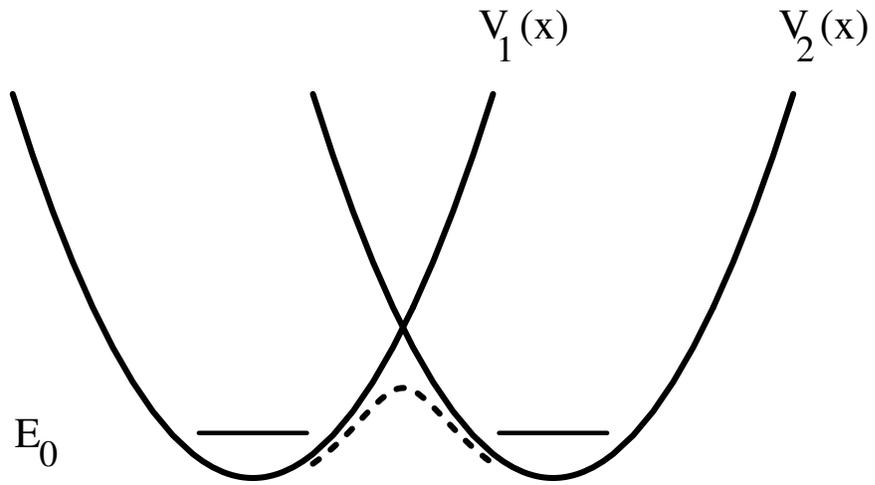
\begin{figure}
\caption{Quantum coherence in the diabatic picture.}
\label{fig4}
\end{figure}
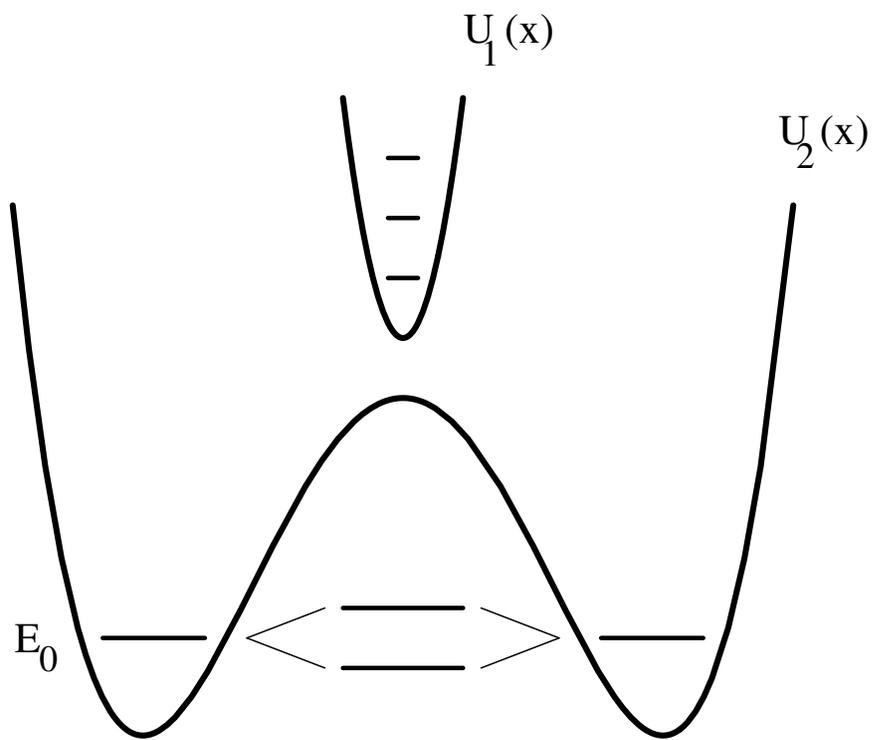
\begin{figure}
\caption{Quantum coherence in the adiabatic picture.}
\label{fig5}
\end{figure}


\end{document}